\input epsf
\documentstyle[preprint,aps]{revtex}

\begin{document}

\count255=\time\divide\count255 by 60 \xdef\hourmin{\number\count255}
  \multiply\count255 by-60\advance\count255 by\time
  \xdef\hourmin{\hourmin:\ifnum\count255<10 0\fi\the\count255}

\draft
\preprint{\vbox{\hbox{UCSD/PTH 00-12}\hbox{JLAB-THY-00-13}
}}

\title{Naturalness of the Coleman-Glashow Mass Relation in the $1/N_c$
Expansion: an Update}

\author{Elizabeth Jenkins$^\ddagger$\footnote{ejenkins@ucsd.edu} and
Richard F. Lebed$^\S$\footnote{lebed@jlab.org}}

\vskip 0.1in

\address{$^\ddagger$Department of Physics, University of California at
San Diego, La Jolla, California 92093\\ $^\S$Jefferson Lab, 12000
Jefferson Avenue, Newport News, VA 23606}

\vskip .1in
\date{May, 2000}
\vskip .1in

\maketitle
\tightenlines
\thispagestyle{empty}

\begin{abstract}
	A new measurement of the $\Xi^0$ mass 
verifies the accuracy of the Coleman-Glashow relation at the level predicted 
by the $1/N_c$ expansion.  Values for other baryon 
isospin mass splittings are updated, and continue to agree with the $1/N_c$
hierarchy.  
\end{abstract}


\newpage
\setcounter{page}{1}

	The recent measurement of the $\Xi^0$ mass 
$1314.82 \pm 0.06 \pm 0.2$ MeV by the NA48
collaboration\cite{NA48} represents a significant improvement over
the 30-year-old value $1314.9 \pm 0.6$ MeV\cite{PDG}.  The $\Xi^0$ mass
now is known to an uncertainty comparable to that of the other baryons of the
lowest-lying spin-$1/2$ octet.  This improvement makes it possible to test
the precision of the famous Coleman-Glashow (CG) mass relation\cite{CG}
\begin{equation} \label{CG1}
\Delta_{\rm CG}= (p-n) - (\Sigma^+ - \Sigma^-) + (\Xi^0 - \Xi^-) = 0.
\end{equation}
Using the old and new experimental values
for $\Xi^0$ yields $\Delta_{\rm CG} = 0.39 \pm 0.61$ and $0.29 \pm
0.26$ MeV, respectively: For the first time, $\Delta_{\rm CG}$ has
been measured to have a nonzero value, though only at the one-sigma level.  
It is of theoretical interest to understand the size of this breaking.  
In this note, we observe that the experimental value agrees with
the theoretical accuracy of the CG relation as predicted in the $1/N_c$ 
expansion of QCD~\cite{JL}.

The mass spectrum of the 
baryon spin-$1/2$ octet and spin-$3/2$ decuplet was analyzed 
in Ref.~\cite{JL} in a combined expansion in  
$1/N_c$ and flavor-symmetry
breaking.  It was found that all of the baryon mass splittings have a natural
explanation in terms of powers of $1/N_c$, $SU(3)$ breaking $\epsilon$, and
isospin breaking $\epsilon^\prime$ (from $m_d - m_u$) or
$\epsilon^{\prime\prime}$ (from electromagnetic effects).  Our analysis differs
from the standard flavor-symmetry breaking analysis in that it incorporates
the enhanced symmetry of baryons present in the large-$N_c$ limit.  Large-$N_c$
baryons respect an exact $SU(6)$ spin-flavor 
symmetry~\cite{DM,J,DJM}.\footnote{For a recent
review, see Ref.~\cite{review} and references therein.}  
For arbitrary $N_c$, the
ground state baryons fill the $N_c$-quark completely symmetric
representation of the spin-flavor algebra, which for $N_c=3$
reduces to the usual ${\bf 56}$-plet of $SU(6)$.  The spin-flavor symmetry is
broken by corrections of subleading order in $1/N_c$, while flavor symmetry
is broken in the usual manner.  Our analysis in Ref.~\cite{JL} showed that the
CG mass combination is $O(\epsilon^\prime \epsilon/N_c^2)$
relative to the average mass of the baryon ${\bf 56}$ spin-flavor multiplet,
which is of order $N_c \Lambda_{\rm QCD}$.  For $N_c=3$, this result implies
that the CG mass combination is predicted to be an order of magnitude smaller 
than expected from an $SU(3)$ flavor symmetry-breaking analysis alone.

In this work, we update the experimental values of mass combinations affected
by the new mass measurement of the $\Xi^0$.  First, we 
briefly review notation introduced in Ref.~\cite{JL}: The isospin
$I$ combinations of baryon masses are denoted by a subscript $I$.
Thus, the $I=0$ and $I=1$ mass combinations of the 
$\Xi^0$ and $\Xi^-$ masses are denoted by
\begin{eqnarray}
\Xi_0 &=& \frac 1 2 (\Xi^0 + \Xi^-) , \nonumber\\ 
\Xi_1 &=& (\Xi^0 - \Xi^-),
\end{eqnarray}
respectively.  Using the new value of the $\Xi^0$ mass changes the experimental
values of these mass combinations to
\begin{eqnarray}
\Xi_0 & = & 1318.07 \pm 0.12 \ ({\rm was } \ 1318.11 \pm 0.31) \ {\rm
MeV}, \nonumber \\ \Xi_1 & = & \ \ -6.50 \pm 0.25 \ ({\rm was } \ \ 
-6.4 \ \pm 0.6) \ \ \ {\rm MeV} .
\end{eqnarray}
The improvement in the experimental value of the $I=0$ mass combination
$\Xi_0$ is small, and does not appreciably affect the numerical evaluation
performed
in Ref.~\cite{JL} of $I=0$ mass combinations.  Thus,
we restrict our attention here to $I=1$ mass combinations.  
The remaining $I=1$ mass
combinations are denoted by  
\begin{eqnarray}
& &
N_1        = (p-n), \ \ \ \ \ \ \ \ \ \,
\Sigma_1   = (\Sigma^+ - \Sigma^-), \ \ \
\Delta_1   = (3 \Delta^{++} + \Delta^+ - \Delta^0 - 3 \Delta^-),
\nonumber \\ & &
\Sigma^*_1 = (\Sigma^{*+} - \Sigma^{*-}), \ \ \
\Xi^*_1    = (\Xi^{*0} - \Xi^{*-}),
\end{eqnarray}
and the $\Lambda$-$\Sigma^0$ mixing parameter.  In terms of these definitions,
the CG mass combination is given by $\Delta_{\rm CG} = N_1 - \Sigma_1 + \Xi_1$.
There are large uncertainties in the $\Delta$ isospin mass splittings, so the
$I=1$ mass combination $\Delta_1$ does not figure in the mass combinations we
consider.

	As in Ref.~\cite{JL}, we define the {\it relative accuracy\/}
$R$ of a linear combination of masses written in the
form $\ell - r$ (where the combinations $\ell$ and $r$ are uniquely
defined to contain baryon masses with only positive coefficients) by $R \equiv |\ell -
r|/[(\ell + r)/2]$.  The quantity $R$ yields a 
scale-independent measure of the
breaking of the relation compared to the average baryon mass.
The theoretical expectation $R_T$ for
$R$ of a particular mass combination 
is given by the combined flavor and $1/N_c$ suppressions of the mass
combination, which are listed in Table~II of Ref.~\cite{JL},
divided by $N_c$ since the average baryon mass is $O(N_c^1)$. 
As an example, $R_T =
\epsilon^\prime \epsilon /N_c^2$ for the CG combination.  If the
$1/N_c$ expansion is natural, then $R/R_T$ should be a number of order
unity.

In Table~I we present values of $R$ and $R_T$
for the four mass combinations depending upon $\Xi_1$ in Table~II
of Ref.~\cite{JL}.
We obtain numerical values for $R_T$ by taking $\epsilon
\approx 1/4$ and $\epsilon^\prime \approx 1/3^5$ (and of course
$N_c=3$); these are typical values one finds for flavor breaking in
the meson mass spectrum, but one could also in principle fit to them
using the observed baryon masses.  One sees first that only the CG
combination central value changes substantially from the improvement
of the $\Xi^0$ mass measurement, although three of the four
uncertainties drop significantly.  Most importantly, one observes that
the combined $1/N_c$ and flavor expansion continues to explain the
size of the mass combinations in a natural way.  It is also clear from
Table~I that the agreement of $R$ and $R_T$ would simply fail without
the explicit $1/N_c$ factors.

	The improvement in the measured value of $\Xi_1$ also permits 
a better estimate of
the $\Lambda$-$\Sigma^0$ mixing parameter.  Using Eq.~(4.10) of
Ref.~\cite{JL}, we find 
\begin{equation}
\Lambda\Sigma^0 = \frac{1}{2\sqrt{3}} ( \Xi_1 - N_1 ) = -1.50 \pm 0.07 \
({\rm was} \ -1.47 \pm 0.17) \ {\rm MeV} ,
\end{equation}
up to a theoretical uncertainty of $O(\epsilon^\prime \epsilon / N_c^2)$
times the average baryon mass, 
which yields a comparable theoretical error.

	In summary, the new $\Xi^0$ mass measurement leads to a one-sigma
determination of the magnitude of the Coleman-Glashow mass combination.
The current experimental value of the CG mass combination is naturally 
explained in the $1/N_c$ expansion, which yields an additional suppression
factor of $1/N_c^2$ beyond flavor symmetry-breaking factors;
an $SU(3)$ flavor symmetry-breaking analysis alone fails to explain the
observed accuracy of the CG mass combination.  Further testing of the 
mass hierarchy predicted in the combined $1/N_c$ and flavor-symmetry breaking
expansion is possible by improving the measurements of isospin mass splittings
in the decuplet, particularly those of the $\Delta$ baryon.

{\samepage
\begin{center}
{\bf Acknowledgments}
\end{center}
EJ was supported by the Department of Energy under Contract No.\
DOE-FG03-97ER40546 and by the National Young Investigator program through Grant
No. PHY-9457911 from the National Science Foundation.
RFL was supported by the Department of Energy under Contract No.\
DE-AC05-84ER40150.}

\newpage

\begin{table}

\begin{tabular}{ccccc}
Mass combination & $R$ (old) & $R$ (new) & $R_T$ & $R_T$ (num) \\
\hline \\
$N_1 - \Sigma_1 + \Xi_1$ & $(1.1 \pm 1.8) \times 10^{-4}$ 
& $(8.4 \pm 7.5) \times 10^{-5}$ & $\epsilon^\prime \epsilon / N_c^2$ 
& $1.1 \times 10^{-4}$ \\

$25(\Sigma_1 + \Xi_1) - 3(4\Sigma^*_1 - 3\Xi^*_1)$ 
& $(3.6 \pm 0.2) \times 10^{-3}$ & $(3.7 \pm 0.1) \times 10^{-3}$ 
& $\epsilon^\prime / N_c$ & $1.4 \times 10^{-3}$ \\

$N_1 - \Xi_1$ & $(2.3 \pm 0.3) \times 10^{-3}$ 
& $(2.3 \pm 0.1) \times 10^{-3}$ & $\epsilon^\prime / N_c$ 
& $1.4 \times 10^{-3}$ \\

$5(2N_1 + \Sigma_1 - \Xi_1) - 3(4\Sigma^*_1 - 3\Xi^*_1)$ 
& $(0.5 \pm 1.8) \times 10^{-4}$ & $(0.6 \pm 1.7) \times 10^{-4}$ 
& $\epsilon^\prime / N_c^3$ & $1.5 \times 10^{-4}$ \\

\end{tabular}
\caption{$I=1$ mass combinations before (old) and after (new) the recent 
measurement of the $\Xi^0$ mass. The relative accuracy $R$ and its 
theoretical value $R_T$ are defined in the text.}
 
\end{table}

\end{document}